# Machine-agnostic Automated Lumbar MRI Segmentation using a Cascaded Model Based on Generative Neurons

Promit Basak[1], Rusab Sarmun[1], Saidul Kabir[1], Israa Al-Hashimi[2], Enamul Hoque Bhuiyan[3], Anwarul Hasan[4], Muhammad Salman Khan[5], Muhammad E. H. Chowdhury[5*]

[1]Department of Electrical and Electronic Engineering, University of Dhaka, Dhaka 1000, Bangladesh. Email: promit-2017614892@eee.du.ac.bd (PB), rusab-2016315002@eee.du.ac.bd (RS), saidul-2016614965@eee.du.ac.bd (SK)

[2]Department of Radiology, Hamad Medical Corporation, Doha, Qatar. Email: Ialhashimi@hamad.qa (IA)

[3]Center for Magnetic Resonance Research, University of Illinois Chicago, IL 60612, USA. Email: bhuiyan@uic.edu (EHB)

[4]Department of Industrial and Mechanical Engineering, Qatar University, Doha-2713, Qatar. Email: ahasan@qu.edu.qa (AH)

[5]Department of Electrical Engineering, Qatar University, Doha 2713, Qatar. Email: salman@qu.edu.qa (MSK), mchowdhury@qu.edu.qa (MEHC)

*Corresponding author: Muhammad E. H. Chowdhury (mchowdhury@qu.edu.qa)

**Abstract**

Automated lumbar spine segmentation is very crucial for modern diagnosis systems. In this study, we introduce a novel machine-agnostic approach for segmenting lumbar vertebrae and intervertebral discs from MRI images, employing a cascaded model that synergizes an ROI detection and a Self-organized Operational Neural Network (Self-ONN)-based encoder-decoder network for segmentation. Addressing the challenge of diverse MRI modalities, our methodology capitalizes on a unique dataset comprising images from 12 scanners and 34 subjects, enhanced through strategic preprocessing and data augmentation techniques. The YOLOv8 medium model excels in ROI extraction, achieving an excellent performance of 0.916 mAP score. Significantly, our Self-ONN-based model, combined with a DenseNet121 encoder, demonstrates excellent performance in lumbar vertebrae and IVD segmentation with a mean Intersection over Union (IoU) of 83.66%, a sensitivity of 91.44%, and Dice Similarity Coefficient (DSC) of 91.03%, as validated through rigorous 10-fold cross-validation. This study not only showcases an effective approach to MRI segmentation in spine-related disorders but also sets the stage for future advancements in automated diagnostic tools, emphasizing the need for further dataset expansion and model refinement for broader clinical applicability.

## 1. Introduction

The spine is regarded as one of the most critical parts of the body, as it is responsible for upright posture, movement, and structural support. Besides its mechanical tasks, the spine protects the spinal cord, a neurological conduit that connects the brain and body. This complicated network of nerves allows voluntary control of physical motions and guarantees correct organ function. As a result, maintaining spinal health becomes a critical requirement for anyone trying to live an active and useful life. Anatomically, the spinal column comprises 33 tiny bones known as "vertebrae" piled on top of each other. A soft, gel-like cushion called an "Inter-Vertebral Disc" (IVD) sits between each vertebra, absorbing pressure and preventing the bones from grinding against each other. The vertebrae are numbered and can be divided into five distinct regions from top to bottom: cervical, thoracic, lumbar, sacrum, and coccyx (Bogduk, 2005). An anatomical structure of the spine is shown in Figure 1.

Lower back pain (LBP) is a prevalent condition that naturally occurs, particularly among adults, leading to restricted movement and impacting daily activities (Hoy, et al., 2014). The high incidence of degenerative spinal disorders, notably among working-age individuals, results in substantial societal costs associated with treatment and disability (Fayssoux, Goldfarb, Vaccaro, & Harrop, 2010). Lumbar spinal stenosis (LSS) is identified as a primary cause of low back pain and one of the leading reasons for spine surgery (Deyo, et al., 2010). When confronted with symptoms of LBP, the initial course of action typically involves investigating spinal structural issues. Various medical imaging techniques, including x-rays, computed tomography (CT), magnetic resonance imaging (MRI), and positron emission tomography (PET), have been widely utilized over the past few decades for diagnosing spinal problems (Jeon & Kong, 2020; Sneath, Khan, & Hutchinson, 2022). Though CT and MRI are both very effective in detecting herniated discs and spinal stenosis for diagnosing infections or malignancies causing back pain, MRI is more sensitive and specific than conventional imaging techniques(Jarvik & Deyo, 2002). Specifically, MRI is considered more reliable for radiographic assessment of degenerative lumbar spinal stenosis(Alsaleh, et al., 2017).

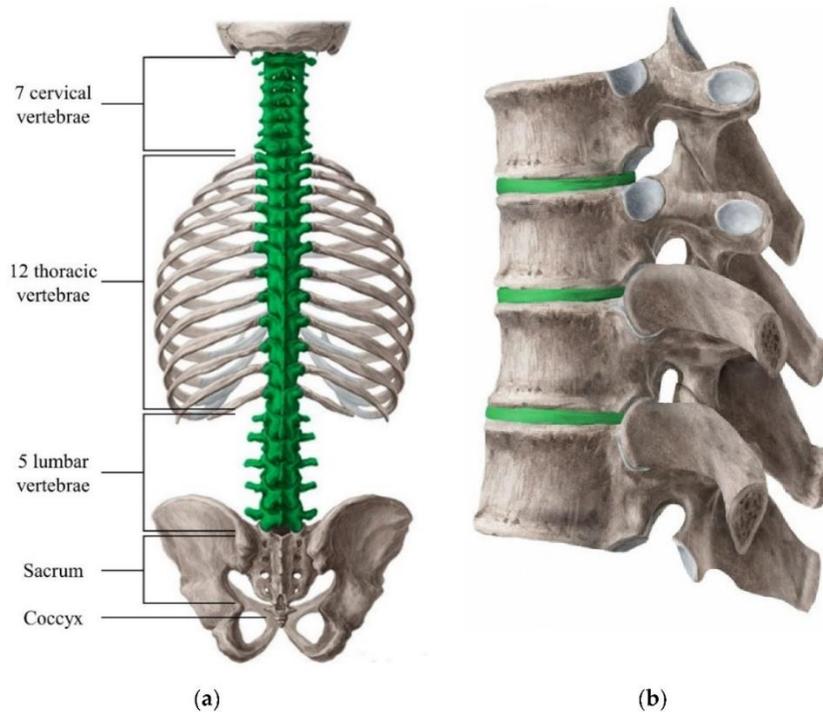

*Figure 1: Overview of the vertebral column. (a)part of the spine that contains vertebrae and IVD (highlighted in green), and (b)the structure of vertebrae and IVD (highlighted in green) (Frost, Camarero-Espinosa, & Foster, 2019).*

Magnetic resonance imaging (MRI) is a noninvasive diagnostic procedure that generates highly detailed pictures of nearly all internal structures within the human body, encompassing organs, bones, muscles, and blood vessels(Westbrook & Talbot, 2018). Unlike X-rays, MRI scanners generate bodily pictures by employing a powerful magnet and radio waves and do not generate ionizing radiation. Though the use of MRI scans can provide the referring clinician with accurate information about the spinal condition, diagnosing lower back pain can be a difficult and time-consuming process compared to X-rays.

Currently, the spinal surgeons and radiologist community rely on either fully manual or semi-automated techniques. They are required to identify each spine segment by analyzing numerous MR images and manually delineate any abnormal regions. For instance, the physician may need to examine MRI scans from various perspectives or use different imaging modalities in order to determine the extent of the lesion. Consequently, the task of analyzing MRI slices is problematic because it requires simultaneously examining and analyzing extensive multimodal data to determine the location and morphology of the lesion. Furthermore, it is seen that the clinical diagnosis of MRI often shows notable differences among observers. This is because experienced radiologists, who rely on their own expertise, generate different radiological grading results even when using the same grading criteria (Lee, et al., 2010). An automated diagnostic system has the capability to address both issues and aid doctors in analyzing MR images efficiently and precisely.

Segmenting MR images has several challenges associated with:
1. Due to the low contrast of MRI images, the boundary between the spine and surrounding tissues is usually indistinct.
2. The shape of the vertebrae or IVDs varies greatly and may fluctuate dramatically across patients.
3. Variation among slices of the same MRI scan is too significant to create a single generalized approach for all distinct slices.
4. Because of technology differences, inter-scanner variability may arise.

5. Intra-scanner variability can occur due to mobility, physiological state, medication, time of day, and other factors (Wittens, et al., 2021).

Finally, the presence of foreign bodies, irrelevant organs, and noise adds another layer of challenge to this task.

To address the problems of spinal MRI interpretation, a range of computer-aided diagnostic strategies have been investigated for possible application throughout the last decade. These applications include computer vision algorithms to segment or locate vertebral bodies and discs, such as histogram-oriented gradients (Ghosh, Malgireddy, Chaudhary, & Dhillon, 2012; Lootus, Kadir, & Zisserman, 2014; A. B. Oktay & Akgul, 2013), probabilistic models (M. S. Aslan, Ali, Rara, & Farag, 2010; Bejnordi, et al., 2017; Corso, Alomari, & Chaudhary, 2008), and GrowCut (Egger, Nimsky, & Chen, 2017). These models mainly localized or detected the target parts and were very complex, time-consuming, and lacked performance. However, the scenario changed after the burgeon of deep learning. Notably, the proposal of U-net (Ronneberger, Fischer, & Brox, 2015) brought about a revolution in medical image segmentation tasks. Some other networks based on U-net, such as Unet++(Z. Zhou, Siddiquee, Tajbakhsh, & Liang, 2019), nnU-net(Isensee, et al., 2018), Attention U-net(O. Oktay, et al., 2018), and TransUnet (J. Chen, et al., 2021) worked very well as a generalized segmentation model. Meanwhile, some other approaches like PSPNet (Zhao, Shi, Qi, Wang, & Jia, 2017), DeepLabV3 (L.-C. Chen, Papandreou, Schroff, & Adam, 2017), and PAN(H. Li, Xiong, An, & Wang, 2018) showed their ability to achieve outstanding performance in segmentation tasks. Hence, deep learning has also become the predominant approach in the spine segmentation domain. A number of research have already been conducted in lumbar MR image segmentation tasks based on deep learning approaches (Han, Wei, Mercado, Leung, & Li, 2018; S. Wang, Jiang, Yang, Li, & Yang, 2022; Whitehead, Moran, Gaonkar, Macyszyn, & Iyer, 2018; Yilizati-Yilihamu, Yang, Yang, Rong, & Feng, 2023; R. Zhang, Xiao, Liu, Li, & Li, 2020) and some of them achieved state-of-the-art performance. However, almost every one of the studies primarily examines MRI data acquired from a single MRI scanner, disregarding the differences between scanners, and so lacks broad application. The majority of the studies only focus on segmenting the entire spine, spinal canals, or discs, which restricts their practical usefulness and potential for further development. Importantly, none of the previous research could attain a result comparable to human performance and can be effectively applied in practice.

In this paper, we propose a novel approach based on a Self-organized Operational Neural Network (Self-ONN) that addresses all the issues stated above to realize an automatic MR image segmentation system that can work in a machine-agnostic way to better generalize across several scanners or patients. The main contributions of our work are listed below:
- We proposed a machine-agnostic approach that provides a generalized outstanding performance on images from a variety of MRI scanners.
- Our proposed system can accurately segment individual lumbar vertebral bodies and intervertebral discs, which is the first of its kind to the best of our knowledge.
- Despite segmenting vertebral bodies and IVDs individually, the proposed framework achieves a comparable performance to the state-of-the-art studies segmenting them non-exclusively.

This paper is divided into five sections. In this section, we provide a brief explanation of the study's motivation and the challenges present in automated MRI segmentation. Section 2 will look further at comparable works and their contributions and limitations. The following Section presents a conceptual framework that underlies our proposed technique. The results of this research are summarized in Section 4. Finally, Section 5 includes the conclusion and future hopes.

## 2. Related Works

Researchers have been working on lumber vertebrae and intervertebral disc segments for quite some time. Nagel et al.(Naegel, 2007) proposed a mathematical morphology-based method for anatomically labeling vertebrae from 3D CT-scan images. Ghosh et al.(Ghosh & Chaudhary, 2014) used a two-stage algorithm to detect the spinal cord with the Hough transform (Deans, 1981) and then extract the intervertebral discs using an adaptive window. A similar approach was proposed by Bhole et al.(Bhole, Kompalli, & Chaudhary, 2009) where they detected the centers of the discs from T1 and T2 weighted sagittal and axial images. Aslan et al.(M. Aslan, Shalaby, Ali, & Farag, 2015) developed a novel probability energy function that incorporates intensity, spatial interaction, and shape information. They then tuned this function to get the best possible segmentation. These studies answered the primary questions and opened the door for further research in this field. However, most of them presented manual or semi-automatic processes that still required expert supervision to operate. The results were also not very promising to be used in real life.

The revolution in computing power made it possible to work with deep neural networks which had a significant impact in lumbar segmentation from MR images as well. Spine-GAN, as proposed by Han et al.(Han, et al., 2018), used two different networks: a generator and a discriminator network to segment different vertebrae and IVDs from T1 and T2 weighted MR images. They achieved a very good pixel-level accuracy of 96.2% and a dice coefficient of 87.1% in their six-class segmentation task. Whitehead et al.(Whitehead, et al., 2018), used four simultaneous FCN networks trained using images at different scales to utilize spatial features at different dimensions. They used MR images from the UCLA PACS database and hand-annotated the lumbar vertebrae and IVDs, but they did not differentiate between different vertebrae or discs. In another study conducted by Zhang et al.(R. Zhang, et al., 2020), an adversarial network was used to localize and segment the vertebral body. They proposed a novel XOR loss for the discriminator and achieved a commending dice score of 95.3%. However, their work concentrated only on segmenting the vertebral body, not the intervertebral discs.

U-net seemed to be a very popular choice in this task as several works were done based on these networks (Dolz, Desrosiers, & Ben Ayed, 2018; Liu, Deng, & Liu, 2021; Lu, Li, Yu, Zhang, & Yu, 2023; S. Wang, et al., 2022; Z. Wang, Xiao, & Tan, 2023; Q. Zhang, et al., 2021). Zhang et al. (Q. Zhang, et al., 2021) aimed to expedite the spine segmentation process using a BN-Unet architecture and claimed to increase the speed more than five times compared to the current Unet architecture. In exchange, they segmented the whole spine as a single instance instead of different parts from T1 and T2 sagittal MR images from 22 subjects. The accuracy and sensitivity obtained were 94.5% and 84.7%, respectively. Wang et al. (S. Wang, et al., 2022) modified the attention U-net by adding attention gates at the end of the encoder and decoder parts. They also changed the down-sampling and residual layers to get a sensitivity of 94% and a dice score of 95%.

Apart from variants of U-net, Yilihamu et al. (Yilizati-Yilihamu, et al., 2023) proposed a scene-aware fusion network (SAFnet) that extracts features at three levels using five residual blocks, and each layer is associated with CBR processing. The different levels of features were then aggregated using a multiscale fusion module. They used a total of 172 MR images and segmented them into 17 classes of vertebrae and IVDs to achieve an average ice coefficient of 79%.

The common lack of all the research was the inapplicability of using their approaches in a generalized experimental setup. Most of the datasets included MR images from a single machine operating in a fixed set of parameters, including repetition time, echo time, flip angle, etc. However, this will not be applicable in practice where the machine and operating setup will be different. Another drawback is the objective, as most of the studies aimed to extract all the vertebrae or IVDs instead of segmenting each of them separately. The performance should also be improved in order to be deployed in real life. In this paper, our goal is to answer all the challenges stated earlier by designing a generalized machine-agnostic framework that can segment different lumbar vertebral bodies and intervertebral discs accurately and efficiently so that it can be deployed in real life.

## 3. Methodology
### 3.1 Overview of the framework

One of the significant properties of the dataset is the annotation is done only on lumbar vertebrae and IVDs even though other parts of the spine, including the thoracic, sacrum, or coccyx, might be present in the scan. To address this issue, we propose a cascaded approach, including a region of interest (ROI) extraction and a segmentation model to efficiently segment different lumbar parts from MR images in a machine-agnostic way. We used a YOLOv8 (Jocher, Chaurasia, & Qiu, 2023) detection model for extraction of the lumbar spine followed by a Self-ONN(Kiranyaz, et al., 2021) based encoder-decoder architecture for the segmentation task. A high-level overview of our used framework is given below:

- The raw MR image slices were extracted into images and underwent extensive preprocessing.
- The manually annotated segmentation masks from different classes were unified into a single mask using the STAPLE algorithm (Warfield, Zou, & Wells, 2004).
- A subject-preserved cross-validation set is generated for a fair evaluation of the proposed pipeline.
- The ROI from the preprocessed images was extracted using the YOLOv8 model.
- The extracted ROIs were segmented into nine different classes using the novel Self-ONN-Unet model.

A flowchart of the overview of our proposed framework is provided in Figure 2.

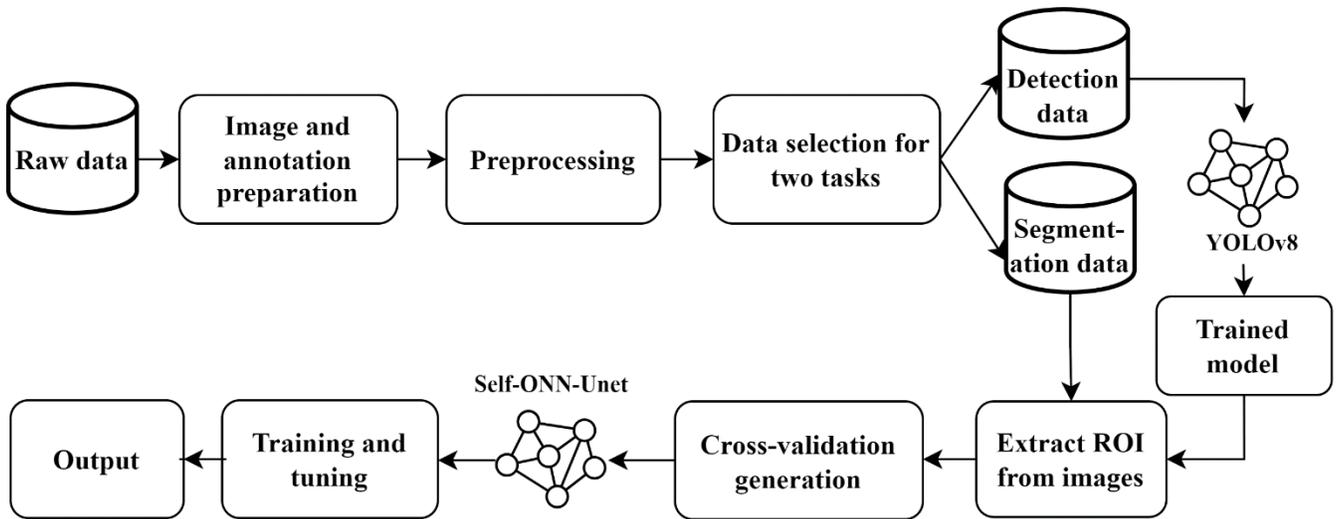

Figure 2: Overview of the proposed framework

*3.2 Preprocessing*

We used a multi-scanner and multi-modal dataset (Khalil, et al., 2022) for this study. This dataset contains MRI scans in different modalities (including T1, T2, STIR, and Dixon) from 12 different scanners. Specific modalities in MRI are customized to emphasize specific tissue properties. T1-weighted MRI has a precise anatomical resolution, making it optimal for viewing typical structures and bleeding. On the other hand, T2-weighted MRI is highly effective in detecting edema and disease due to its heightened sensitivity to water content. STIR (Short Tau Inversion Recovery) is mostly used to suppress fat signals in order to enhance the visibility of fluid and inflammation. The Dixon approach facilitates the differentiation of fat and water signals. Because of their characteristic usage, corresponding images also have different properties. For example, fat appears bright, water and cerebrospinal fluid (CSF) appear dark in T1-weighted MRI, while water and CSF appear bright and fat appear darker in T2-weighted MRI. The difference between various modalities is illustrated in Figure 3:

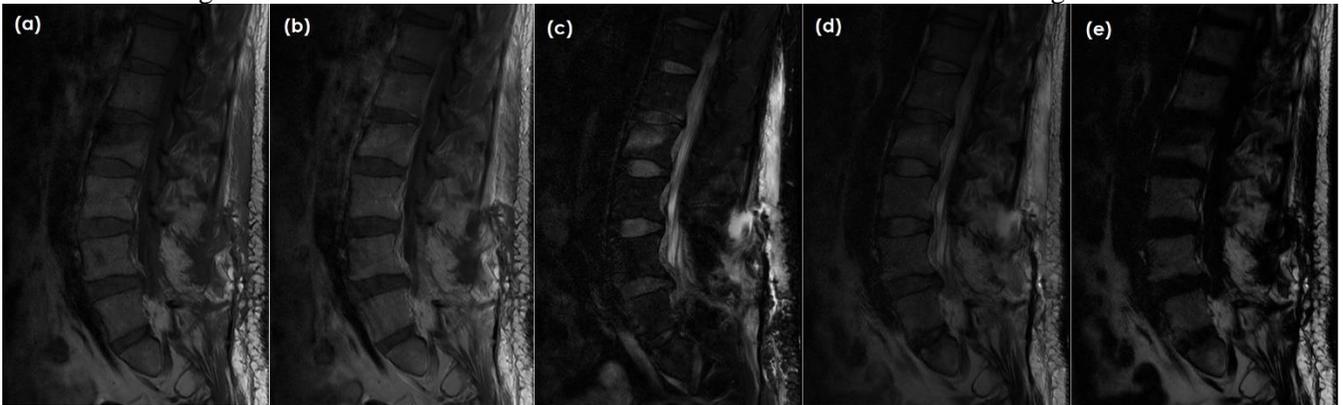

Figure 3: Variability among different types of scans: (a) T1 weighted non-contrast-enhanced, (b) T1 weighted contrast-enhanced, (c) T2 weighted Dixon water enhanced, (d) T2 weighted Dixon, (e) T2 weighted Dixon fat enhanced. All images are taken from the 6$^{th}$ slice of MRI scans of the same patient using a Philips Achieva scanner.

Because of these differences in the image characteristics among different modalities, the model needs a considerable number of scans to learn these properties. Nevertheless, T2, STIR, and Dixon modalities had very few scans available in the dataset. Hence, we used only the T1-weighted MRI for the vertebral body and intervertebral disc segmentation. However, we utilized other modalities in the ROI extraction stage in our cascaded framework.

A common issue regarding volumetric imaging is the variability across different slices. In most cases, the first and last few slices do not or barely contain information about the target tissue or organ, as shown in Figure 4. While this is completely a typical scenario, this occurrence needs careful attention while using deep learning systems.

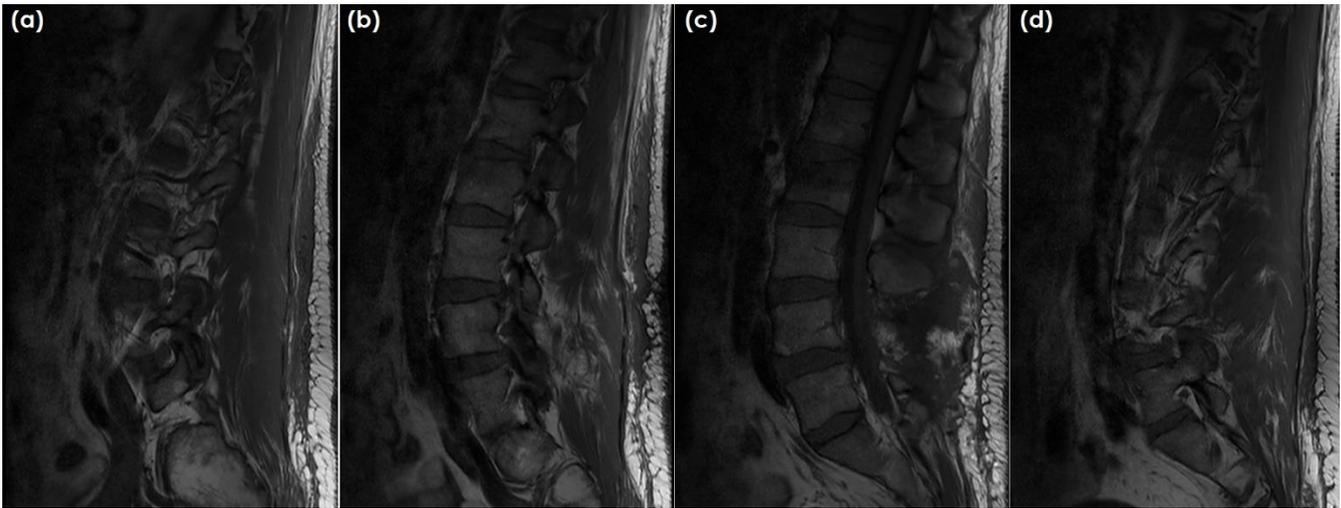

*Figure 4: Inter-slice variability, (a) $1^{st}$ slice, (b) $4^{th}$ slice, (c) $8^{th}$ slice, (d) $15^{th}$ slice of and T1 weighted non-contrast scan using a Philips Achieva scanner.*

As the figure shows, the spine is visible in most of the slices except for the first and last slices. As we inspected different MRI scans, we found the first and last few slices usually do not have a clear view of the spine. Hence, we removed 20% of slices from the beginning and end to remove ambiguous images. The value of 20% was chosen by heuristically investigating the dataset. Still, there might be other noisy images, and it is the responsibility of the ROI extractor model to detect and remove those images using the confidence threshold and the detected non-existent ROI. We also applied image processing techniques to enhance the color and contrast as well as the differentiability of the images. At first, a soft Gaussian blur was applied to the images to remove noise, followed by CLAHE (Reza, 2004) to improve the equalization of the histogram of the MR images.

After modality selection and slice removal process, the number of samples became too small. So, we opted for data augmentation in both stages of the pipeline. However, the augmentation types were different depending on the related task. For the ROI extractor, both geometric (scale, translate, mosaic) and non-geometric (median blur, equalization) augmentations were used, whereas only geometric (translate, rotate, shear) augmentations were applied for segmentation. A summary of our preprocessing algorithm is given in Figure 5:

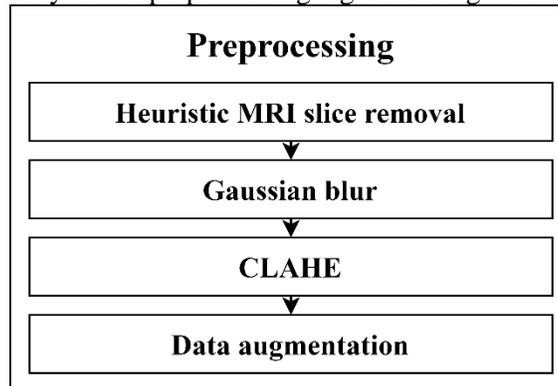

*Figure 5: MRI preprocessing techniques used in this study.*

*3.3 ROI Extraction*

We used the YOLOv8 model for the ROI extraction task. YOLO (You Only Look Once) was originally proposed by Redmon et al.(Redmon, Divvala, Girshick, & Farhadi, 2016). YOLO employed a distinct methodology to utilize a single neural network to process the entire image, bringing revolutionary changes in the object detection field. There has been a collaborative improvement of this model by different researchers, YOLOv8 being one of their latest additions. The main improvement of YOLOv8 is the transition from a traditional anchor-based approach to an anchorless detection system. Anchors solved a few major problems in object detection, including the detection of same-center objects and improving the bounding box area. The architecture of YOLOv8 can be divided into two major parts: backbone and head. The backbone is a modified darknet architecture that extracts information from the image at different spatial levels, and cross-stage partial bottlenecks are added to enhance the information flow between layers. The head consists of 'Conv' (2D convolutional block with Sigmoid Linear Unit (SiLU) activation (Elfwing, Uchibe, & Doya, 2018)) and 'C2f' (partial bottlenecks with two convolutional layers) blocks to predict bounding boxes and class probabilities. In this study, we used the medium-depth model of YOLOv8 (YOLOv8m) with almost 25.9 million parameters to extract the lumbar part of the spine from the MR image. A simplified architecture of YOLOv8 is presented in Figure 6:

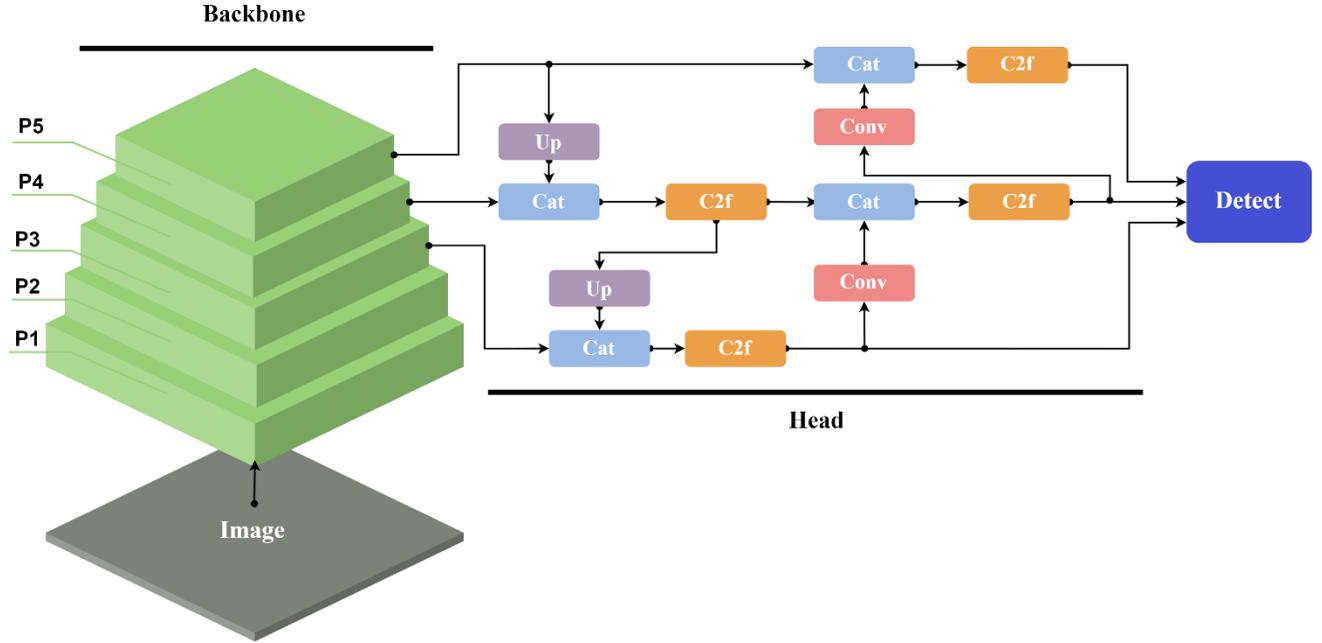

*Figure 6: A simplified architecture of YOLOv8 (Up Upsampling layer, Cat: Concatenation, Conv: single convolution block, C2f: double convolution block with partial bottlenecks).*

*3.4 The notion of Self-ONN:*

Both Multi-Layer Perceptrons (MLPs) and Convolutional brain Networks (CNNs) have a common drawback: they depend on a homogeneous network structure with linear neuron models, which does not accurately capture the variety and intricacy of biological neural systems. Mathematically, the output of the $k^{th}$ neuron of the $l^{th}$ layer of CNN, denoted by $x_k^l$ can be defined by:

$$x_k^l = b_k^l + \sum_{i=0}^{N_{l-1}} x_{ik}^l \qquad (1)$$

where, $b_k^l$ is the bias corresponding to the neuron and $x_i^l$ is the output of $i^{th}$ neuron of $l^{th}$ layer of the CNN. The Operational Neural Networks (ONNs) (Kiranyaz, et al., 2021) tackle this issue by implementing diverse and complex network models including a varied range of operators per neuron, resulting in a more adaptable framework. ONNs' core principle goes beyond relying just on linear convolutions within convolutional neurons. They do this by including nodal and pool operators. These additions include the operational layers and neurons, while still maintaining two fundamental limitations inherited from traditional CNNs: weight sharing and restricted (kernel-wise) connection. Figure 7 depicts the three operational layers and the $k^{th}$ neuron with $3 \times 3$ kernels in an ONN. The input map of the $k^{th}$ neuron in the current layer of an ONN represented as $x_{ik}^l$, is obtained by pooling ($P_k^l$) the final output mappings, , $y_i^{l-1}$, of the neurons in the previous layer that have been processed with their corresponding kernels, $w_{ik}^l$:

$$x_{ik}^l(m) = P_k^l \left( \psi_k^l \left( w_{ik}^l(r), y_i^{l-1}(m+r) \right) \right)_{r=0}^{k-1} \qquad (2)$$

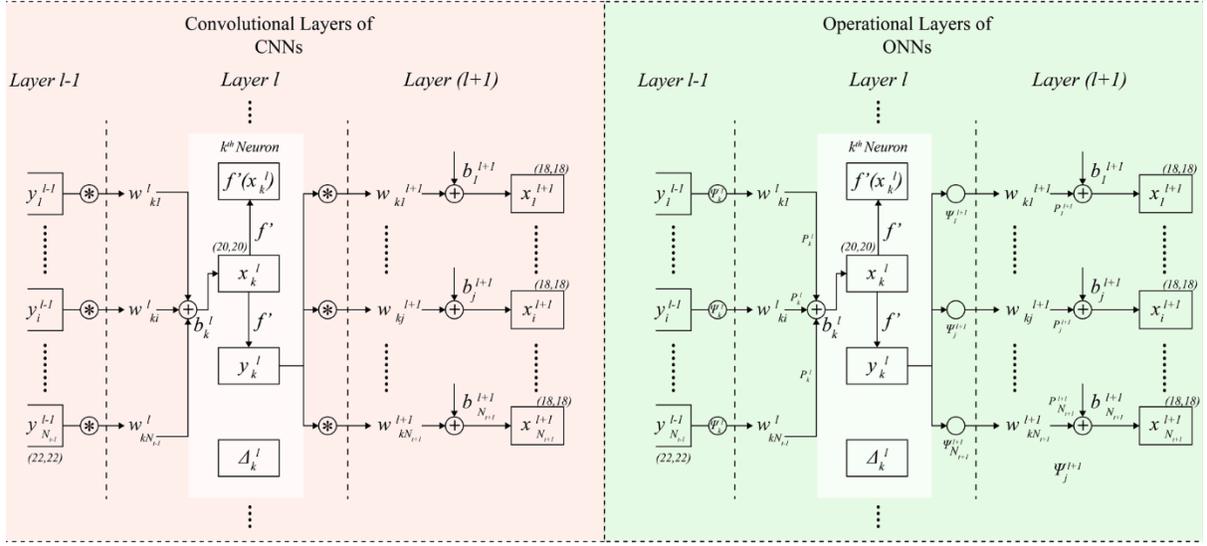

*Figure 7: The illustration of calculating the $k^{th}$ neuron of a CNN (left) and an ONN (right)*

ONNs still have some limitations as they are constrained by the limited number of operators which has a predetermined quantity of manually designed nodal operators. Self-organized Operational Neural Networks (Self-ONNs) expand on the concepts of ONNs by including generative neurons that have the ability to adjust and enhance the nodal operator of each link while undergoing training thus overcoming the limitations of ONNs. Hence, Self-ONNs do not necessitate a previously determined collection of operators or an iterative search procedure to find the most optimal operators. Instead, each neuron has the ability to generate various combinations of nodal operators, which enables a more adaptable and potent modeling capacity. This is accomplished by subjecting every generative neuron of a Self-ONN model to use a composite nodal operator that can be iteratively created during backpropagation training without any restrictions. The composite operator can be represented as a Taylor approximation of order Q, where Q represents the q-order of the polynomial. For instance, the composite nodal operator of the $k^{th}$ generative neuron in the $l^{th}$ layer can be expressed as:

$$\psi_k^l\left(w_{ik}^{l(Q)}(r), y_i^{l-1}(m+r)\right) = \sum_{q=1}^{Q} w_{ik}^{l(Q)}(r,q)\left(y_i^{l-1}(m+r)\right)^q \tag{3}$$

where, $w_{ik}^{l(Q)}$ is the $K \times Q$ dimensional kernel matrix between the $i^{th}$ neuron of the $(l-1)^{th}$ layer to the $k^{th}$ neuron at the current ($l^{th}$) layer, $y_i^{l-1}$ is the output map of the $i^{th}$ neuron at the $(l-1)^{th}$ layer. Now, if the contribution of the $i^{th}$ neuron in generating the feature map from the $(l-1)^{th}$ layer to the $l^{th}$ layer is termed as $x_{ik}^l$, Equation (3) can be alternatively expressed and simplified as,

$$x_{ik}^l = \sum_{q=1}^{Q} Conv\left(w_{ik}^{l(Q)}, \left(y_i^{l-1}\right)^q\right) \tag{4}$$

Utilizing Equation (4), this procedure can be executed by performing convolution operations Q times. Ultimately, the formulation for the output of a single neuron can be expressed as:

$$x_k^l = b_k^l + \sum_{i=0}^{N_{l-1}} x_{ik}^l \tag{5}$$

We can even use Self-ONN like a conventional CNN by setting $Q = 1$ which makes Self-ONN a more generalized version of CNN.

*3.5 Segmentation Model:*

We propose a novel Self-ONN-based encoder-decoder network as the segmentation model of our system. A DenseNet121(Huang, Liu, Van Der Maaten, & Weinberger, 2017) architecture is used as the encoder which is responsible for extracting the features from the image and Self-ONN-Unet as the decoder to generate the segmentation masks.

In DenseNet, layers are densely connected to all the previous layers through dense blocks. As a result, any layer receives a concatenation of the feature maps of all its preceding layers. DenseNet is comprised of three distinct types of blocks. The initial block is the convolution block, which consists of a $3 \times 3$ and a $1 \times 1$ convolution as well as a skip connection to concatenate the output of the previous layer. The second one is the dense block, which is composed of concatenated and densely connected convolution blocks. DenseNet's characteristic component is the dense block. The final block is known as a transition block that links two contiguous dense blocks. Due to the uniformity of feature map proportions within the dense block, the transition layer performs a dimension reduction on the feature map.

In the Self-ONN-based decoder, each Self-ONN block is composed of a Self-ONN layer as the key component, a max pooling layer to select the most important features, and a tanh activation function to add non-linearity. If we go deeper, the Self-ONN layer consists of a number of generative neurons and a dropout layer. There are a total of 5 Self-ONN blocks in our proposed architecture and each of them is connected to the corresponding encoder block with a skip connection. Unlike the traditional Unet architecture, we did not use any bottleneck layer between the encoder and the decoder as the DenseNet121 itself is very deep. A visual representation of our proposed architecture is provided in Figure 8.

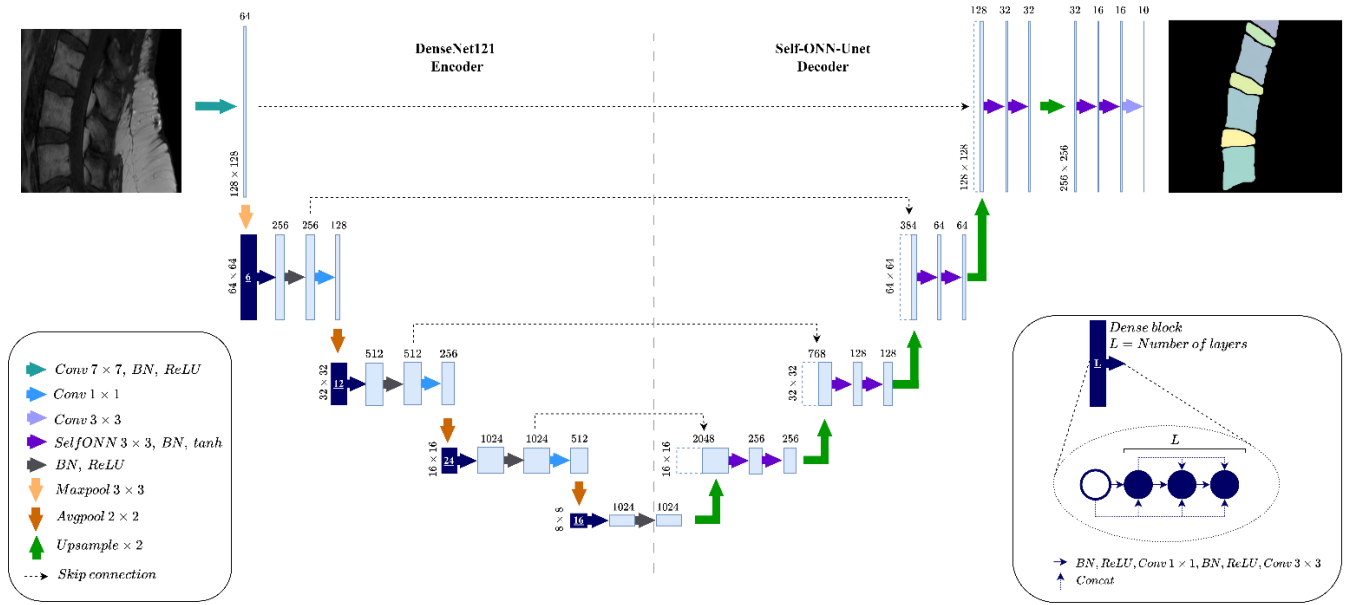

*Figure 8: The detailed architecture of the proposed segmentation model.*

## 4. Result

*4.1 Dataset description*

The "Multi-scanner and multi-modal lumbar vertebral body and intervertebral disc segmentation database" (Khalil, et al., 2022) stands out as one of the few publicly available datasets that integrates data from a variety of machines and modalities to capture lumbar MRI data. This dataset comprises information from 12 distinct scanners and encompasses eight different types of images collected from 34 patients (mean age: 60.4 ± 15.2 years, age range: 30.0–88.1 years, 58.8% females). The patients were categorized based on five clinical indications: 1) lower back pain (LBP), 2) follow-up imaging after resection of a spinal tumor, 3) spinal metastases, 4) spondylodiscitis, and 5) spinal fracture. Only sagittal images covering the lumbar spine were included in the dataset, and the characteristics and parameters of the scanners (including repetition time (TR)/echo time (TE), field of view, and acquisition time) varied across the different types of scanners. Following MR image acquisition, manual segmentation was conducted in the sagittal plane by a medical professional for each lumbar vertebral body (L1, L2, L3, L4, L5) and intervertebral disc (L1-2, L2-3, L3-4, L4-5). An exemplar segmented image from the dataset is illustrated in Figure 9.

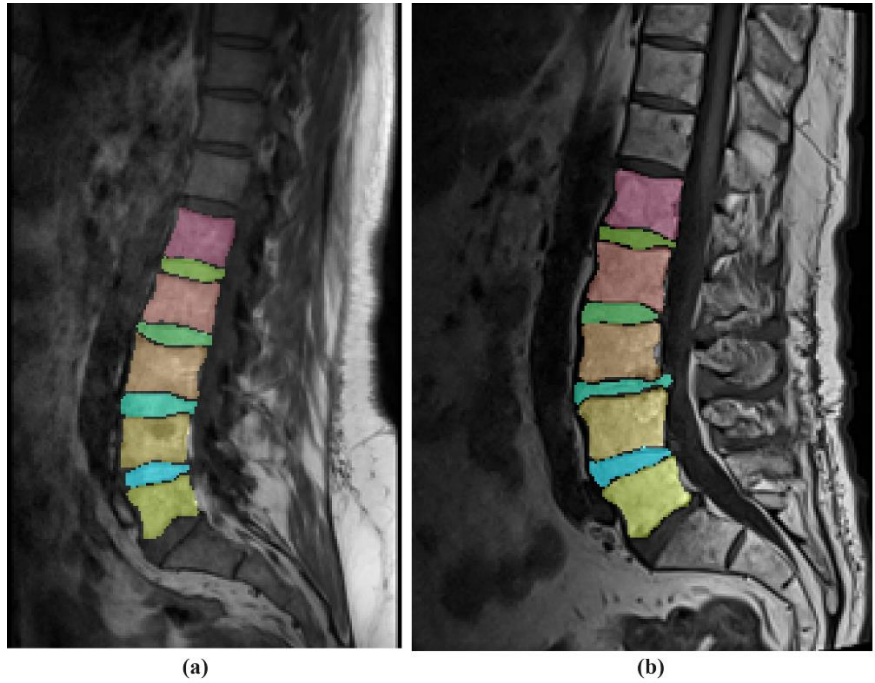

*Figure 9: Segmented images from the dataset. (a) captured using a Philips Achieva scanner, and (b) captured using a Siemens Amira scanner. The segmentation masks are shown in the overlay.*

Unfortunately, corrupted data was present in the dataset. 53 MRI scans out of 211 were corrupted, which comprises almost 25% of the data. A statistical summary of the dataset after removing the corrupted data is presented in Table 1:

*Table 1: Overall statistics of the dataset*

| Number of scans | 156 |
|---|---|
| Number of subjects | 34 |
| Mean age | 60.4 |
| Number of scanners | 12 |
| Number of scan types | 8 |
| Number of slices | 2692 |

The dataset contained 8 different types of MR images including T1 (contrast and non-contrast), T2, STIR, and Dixon (T2, fat and water-saturated) scans. The number of MRI scans and the number of subjects against each scan type is shown in Table 2. It should be noted that each subject can be scanned multiple times on multiple occasions.

*Table 2: Data statistics for each type of scan*

| Scan type | Number of scans | Number of subjects |
|---|---|---|
| T1 non-contrast | 54 | 32 |
| T1 contrast-enhanced | 27 | 21 |
| T1 fat-saturated | 1 | 1 |
| T2 | 7 | 7 |
| STIR | 3 | 3 |
| Dixon T2 | 22 | 19 |
| Dixon Fat | 21 | 18 |
| Dixon Water | 21 | 19 |

*4.2 Dataset preparation:*

  *ROI extraction*: We opted for a rather simple train-validation-test split to train the ROI extraction part. As we had a total of 34 subjects' data, we randomly selected four subjects for the ROI extraction task, and the rest of the 30 subjects' data were kept for the segmentation task. Two of the four chosen subjects were used in the training set, and one was used for the validation and test sets. For the low amount of data, we selected all types of scans including

T1, T2, and Dixon, and all the slices from a single scan for the ROI extraction task. Due to the various types of scans and scanners, the number of scans and slices was not uniform. An overview of the prepared dataset is given in Table 3.

*Segmentation:* For the segmentation model, we selected only T1 scans from different scanners and applied heuristic MRI slice removal to remove slices that do not contain informative images. Later, we applied the ROI extractor to extract the lumbar part of the spine from the whole MRI slice before preparing a subject-wise 10-fold cross-validated dataset. One goal for this study is to make our proposed model learn the variability of different subjects, and machines perform well in unseen test datasets no matter which subject or scanner the data came from. The 30 subjects' data kept for segmentation were used to create a 10-fold cross-validation so that no subject would be present in different folds. Therefore, we ensured the integrity of our model to generalize data from different modalities and prevent data leaking. A summary of the amount of data is provided in Table 3. As the segmentation dataset did not use all the scan types and applied heuristic MRI slice removal, the number of scans and slices was reduced.

*Table 3: Summary of the data split for the cascaded task*

| Task | Split | Number of subjects | Number of distinct scanners | Number of selected scans | Number of selected slices |
|---|---|---|---|---|---|
| ROI Extraction | Train | 2 | 3 | 9 | 160 |
| | Test | 1 | 2 | 6 | 104 |
| | Validation | 1 | 2 | 9 | 147 |
| | **Total** | **4** | **5** | **24** | **411** |
| Segmentation | 1 | 3 | 3 | 9 | 97 |
| | 2 | 3 | 4 | 8 | 96 |
| | 3 | 3 | 3 | 9 | 96 |
| | 4 | 3 | 2 | 6 | 68 |
| | 5 | 3 | 4 | 6 | 69 |
| | 6 | 3 | 3 | 6 | 71 |
| | 7 | 3 | 1 | 5 | 57 |
| | 8 | 3 | 3 | 6 | 64 |
| | 9 | 3 | 5 | 10 | 131 |
| | 10 | 3 | 5 | 8 | 89 |
| | **Total** | **30** | **12** | **73** | **838** |

*4.3 Evaluation metrics*

Evaluation metrics are crucial for measuring the performance of deep learning models and facilitating comparisons across different applications. We employed a range of commonly utilized detection and segmentation metrics to assess the performance of our models across different experimental levels.

*Accuracy (Acc)*: Accuracy quantifies the model's overall capacity to predict the correct output. This metric represents the ratio of correctly predicted observations to the total number of observations. Denoting true positives, false positives, true negatives, and false negatives as TP, FP, TN, and FN, respectively, accuracy can be defined as:

$$\text{Accuracy}, Acc = \frac{TP+FP}{TP+FP+TN+FN} \tag{9}$$

*Precision (Pr)*: Precision quantifies the proportion of correct positive predictions out of all positive predictions, thereby providing useful insights into the model's ability to reduce false positives.

$$\text{Precision}, Pr = \frac{TP}{TP+FP} \tag{10}$$

*Sensitivity (Se)*: Sensitivity, sometimes referred to as recall or true positive rate, refers to a model's ability to completely identify all positive cases in a given dataset.

$$\text{Sensitivity}, Se = \frac{TP}{TP+FN} \tag{11}$$

*Intersection over Union (IoU):* Intersection over Union (IoU) calculates the amount of overlap between the predicted bounding box or segmented region and the ground truth bounding box or annotated. Mathematically, IoU is computed by dividing the intersection of the predicted and ground truth areas by their union, expressed as:

$$IoU = \frac{Area\ of\ intersection}{Area\ of\ union} \tag{12}$$

*Mean average precision (mAP):* Mean average precision (mAP) offers a consolidated estimate of a model's performance across different levels of precision and recall. To comprehend the nature of this trade-off, Precision-Recall curves are usually plotted, and average precision (AP) for a single class is calculated by the area under this precision-recall curve. If average precision is denoted by $AP$ and the total number of classes is $N$, it can be shown mathematically,

$$\text{Mean average precision}, mAP = \frac{1}{N}\sum_{i=1}^{n} AP_i \quad (13)$$

Usually, $mAP$ is calculated at a fixed IoU threshold and for different thresholds, the calculated $mAP$ will be different which makes it harder to compare between $mAP$s with different thresholds. A standard procedure is to calculate the $mAP$ at different thresholds from 0.5 to 0.95 with an increment of 0.05 and take the average of all the calculated $mAPs$. This is known as $mAP 0.5: 0.95$.

$$mAP(0.5:0.95) = \frac{1}{10} \times \sum_{t=0.5}^{0.95} mAP:t$$

*Dice similarity coefficient (DSC)*: The Dice Similarity Coefficient, often known as the Dice score, measures the similarity between two sets that overlap. It can be calculated using the following formula:

$$DSC = \frac{2 \times |X \cap Y|}{|X| \cup |Y|} \quad (14)$$

*4.4 Experimental setup*

The experiments in this study were conducted using a cloud-based computing environment equipped with high-performance GPUs. This choice of environment provided us with the computational resources needed to efficiently train and evaluate our deep-learning model on a sizable dataset of pediatric respiratory sounds. Technical specifications of the computing setup are given in Table 4:

*Table 4: Technical specifications of the computing system*

| Component | Specification |
|---|---|
| CPU | Dual-core Intel Xeon Processor (2.2 GHz) |
| GPU | NVIDIA Tesla T4 GPU (16 GB VRAM) |
| RAM | 25 GB |
| Storage | 120 GB SSD |
| Operating System | Ubuntu 22.04 LTS |
| Software Environment | Python 3.9, Torch 1.13.1, CUDA 11.1 |

*4.5 ROI Extraction performance:*

The result of ROI extraction for different variants of the YOLOv8 network is provided below in Table 5:

*Table 5: ROI extraction performance of different models*

| Model | Reference | Precision | Sensitivity | mAP:0.5 | mAP:0.5:0.95 |
|---|---|---|---|---|---|
| YOLOv5n | (Jocher, 2022) | 0.877 | 0.89 | 0.939 | 0.692 |
| YOLOv5m | (Jocher, 2022) | 0.894 | 0.888 | 0.943 | 0.712 |
| YOLOv6n | (C. Li, et al., 2022) | 0.901 | 0.917 | 0.946 | 0.75 |
| YOLOv6m | (C. Li, et al., 2022) | 0.916 | 0.917 | 0.959 | 0.772 |
| YOLOv6 v3.0s | (C. Li, et al., 2023) | 0.968 | 0.845 | 0.93 | 0.789 |
| YOLOv6 v3.0m | (C. Li, et al., 2023) | 0.971 | 0.887 | 0.959 | 0.815 |
| YOLOv6 v3.0l | (C. Li, et al., 2023) | 0.971 | 0.882 | 0.956 | 0.8 |
| YOLO-NAS-s | (deci.ai, 2023) | 0.93 | 0.915 | 0.966 | 0.822 |
| YOLO-NAS-m | (deci.ai, 2023) | 0.946 | 0.92 | 0.982 | 0.855 |
| YOLOv8n | (Jocher, et al., 2023) | 0.946 | 0.97 | 0.977 | 0.809 |
| **YOLOv8m** | **(Jocher, et al., 2023)** | **0.991** | **0.965** | **0.993** | **0.916** |
| YOLOv8l | (Jocher, et al., 2023) | 0.97 | 0.953 | 0.978 | 0.802 |
| YOLOv8x | (Jocher, et al., 2023) | 0.965 | 0.941 | 0.966 | 0.78 |

The YOLOv8-medium model (with a depth multiplier of 0.67 and width multiplier of 0.75) achieved the best result with $mAP(0.5: 0.95)$ of 0.916, $mAP: 0.5$ of 0.993 which signifies the ability of the model to accurately extract the lumbar spine even in higher threshold requirements. Three sample predictions of the ROI extraction system are given in Figure 10.

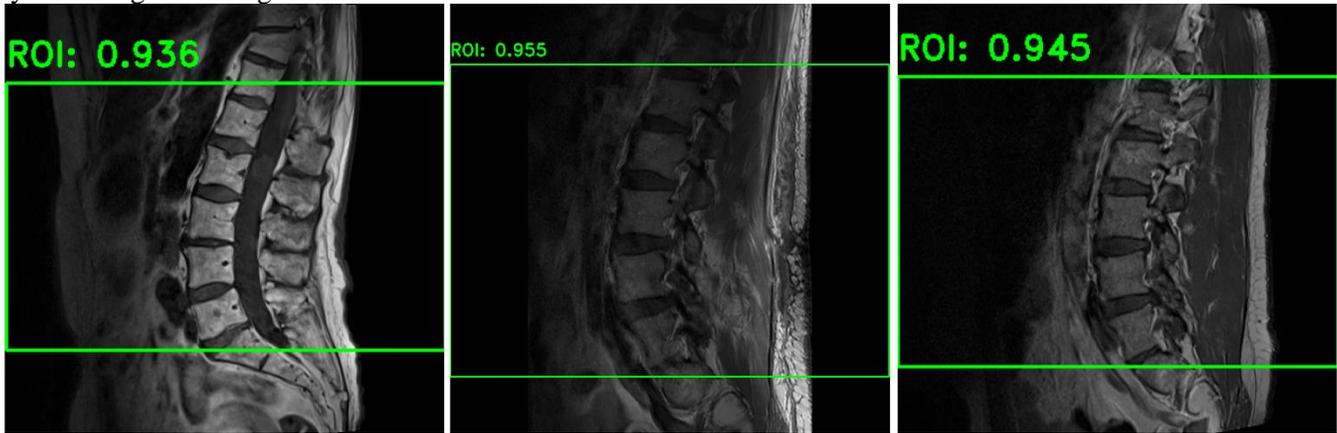

*Figure 10: Sample predictions of the ROI extractor part of the proposed framework.*

*4.6 Segmentation performance:*

Our proposed Self-ONN-based decoder produced an excellent performance on the lumbar vertebra and IVD segmentation compared to other state-of-the-art models we tried. The average result of the 10-fold cross-validation is given below for different architectures in Table 6:

*Table 6: Segmentation performance of different models (metrics are average of 10 folds)*

| Encoder | Decoder | Precision | Sensitivity | Accuracy | Mean IoU | DSC |
|---|---|---|---|---|---|---|
| DenseNet121 | Unet | 86.98±1.2 | 90.06±1.3 | 99.8±0.1 | 79.28±1.9 | 88.33±1.8 |
| ResNet50 | Unet | 86.22±1.2 | 89.66±1.4 | 98.3±0.2 | 78.32±2.3 | 87.57±2.0 |
| Efficientnet-b1 | Unet | 84.66±1.1 | 88.53±1.3 | 98.07±0.1 | 76.4±2.1 | 86.09±1.9 |
| DenseNet121 | Unet++ | 85.89±1.1 | 91.57±0.9 | 98.88±0.2 | 78.18±1.8 | 87.68±1.9 |
| DenseNet121 | Self-ONN-ResUnet | 84.99±1.0 | 86.31±1.1 | 98.54±0.2 | 74.9±2.3 | 85.41±2.1 |
| **DenseNet121** | **Self-ONN-Unet** | **90.71±0.9** | **91.44±1.1** | **99.78±0.1** | **83.66±1.6** | **91.03±1.2** |

Our proposed Self-ONN-Unet decoder combined with the DenseNet121 encoder achieved the best performance among others, with a mean IoU score of 83.657% and dice score of 91.032%. The qualitative result is shown in Figure 11. Detailed qualitative results from multiple patients are shown in Supplementary Figure 1S- 3S.

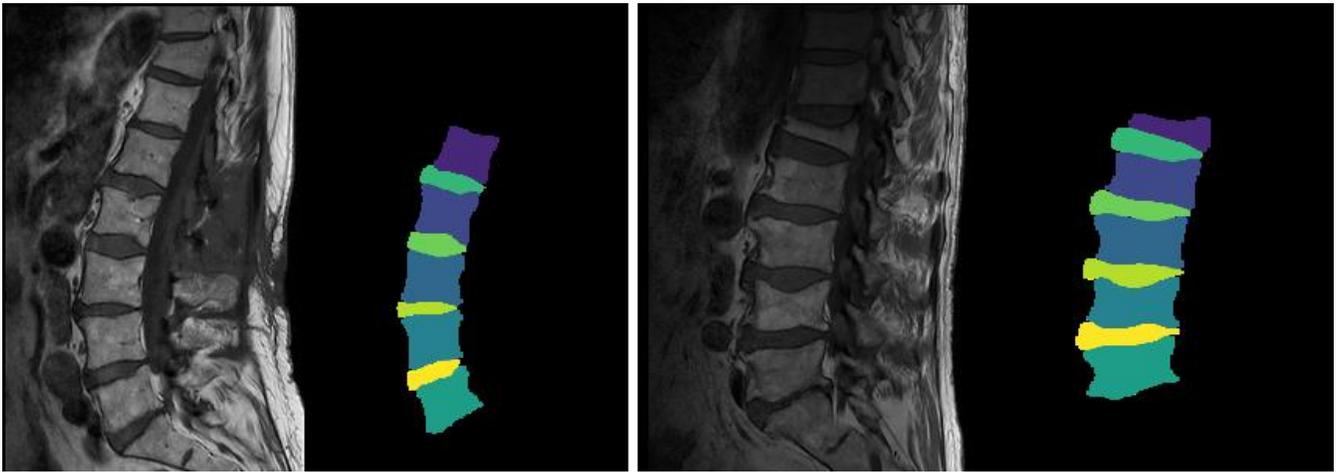
*Figure 11: Two sample predicted segmentations taken from a Philips Ingenia (left) and Philips Achieva scanner (right).*

A comparison of our proposed model with other state-of-the-art approaches in MRI image segmentation is discussed in Table 7:

*Table 7: Comparison with similar state-of-the-art-approaches*

| Year | Reference | Scanners | Classes | Method | DSC | Sensitivity | Other metrics |
|---|---|---|---|---|---|---|---|
| 2018 | (Whitehead, et al., 2018) | 5 | 2: vertebrae, disc | FCN, CNN | Vertebrae: 86.5%, Disc: 83.2% | | |
| 2018 | (Han, et al., 2018) | Multiple | 6: lumber IVD, other IVD, lumbar vertebrae, other vertebrae, NF, NFS | GAN | DSC: 87.1% | 86% | Specificity: 89.1% |
| 2020 | (R. Zhang, et al., 2020) | Not mentioned | 8: L1-L5, S1, T11, T12 (only vertebrae) | Adversarial Network, LSTM | 95% | | |
| 2020 | (J. Zhou, et al., 2020) | 1 | 1: vertebrae | U-net | Rater A: 83.8%, Rater B: 84.9%, | | Rater A: IoU: 75.7% Rater B: IoU: 74.7% |
| 2021 | (Q. Zhang, et al., 2021) | Not mentioned | 1: Whole spine | BN-Unet | | 84.76% | Accuracy: 94.54% Specificity: 86.27% |
| 2022 | (S. Wang, et al., 2022) | Not mentioned | 3: lumber vertebrae, IVD, sacrum | Attention-Unet | 95.01% | 94.53% | Precision: 95.50% |
| 2023 | (Yilizati-Yilihamu, et al., 2023) | Not mentioned | 19: 10 vertebrae, 9 IVDs | SAFNet | 79.46% | | |
| 2024 | (Kim, Park, Lee, & Lee, 2024) | 12 | 1: all the IVDs | CycleGAN | 92% | 90.2% | Mean IoU: 85.3% Precision: 94% |
| | **Proposed** | **12** | **9: 5 vertebrae, 4 IVDs** | **Self-ONN-Unet, DenseNet** | **91.032%** | **91.44%** | **Mean IoU: 83.657%** |

|  |  |  |  |  |  |  | **Accuracy: 99.776%** |

*FCN: Fully Convolutional Network, CNN: Convolutional Neural Network, GAN: Generative Adversarial Network, LSTM: Long Short-Term Memory, SAFnet: Scene Aware Fusion Network, ONN: Operational Neural Network, IVD: Inter-vertebral Disc, NF: Normal Foramen, NFS: Neural Foraminal Stenosis.*

Since our approach to segment each lumbar vertebral body and IVD individually is the first one to the best of our knowledge, it is hard to compare it with similar studies. In addition, most of the studies used a private dataset and did not mention all the related metrics, which makes it even harder to compare. Nevertheless, our proposed method performed similarly or even better compared to a few studies that only segmented the whole spine or all vertebral bodies and IVDs as a single task. (Whitehead, et al., 2018) achieved a DSC of 86.5% on vertebrae and 83.2% on discs, (Han, et al., 2018) achieved a DSC of 87.1% and sensitivity of 86%, (J. Zhou, et al., 2020) achieved a mean DSC of 84.35% across two raters, (Q. Zhang, et al., 2021) achieved a sensitivity of 84.76%, all of which are lower than our achieved DSC of 91.03% and sensitivity of 91.44% though our proposed system segments each part individually. (R. Zhang, et al., 2020) achieved a DSC of 95%, but they only segmented the vertebrae, not the IVDs. (S. Wang, et al., 2022) They also achieved a better DSC (95.01%) but did not segment the vertebrae and IVDs separately. (Yilizati-Yilihamu, et al., 2023) had segmented a large number of vertebrae and IVDs individually, but their obtained DSC (79.46%) is pretty much lower compared to our result. Finally, the only study that used the same dataset similar to ours is (Kim, et al., 2024). They only segmented the whole IVD, but their achieved DSC and sensitivity are very close compared to the performance of the proposed system. Therefore, it is difficult to compare our proposed framework to similar state-of-the-art methods, our study offers excellent performance in this field.

*4.7 Ablation study:*

While designing our model, we came up with different ideas and concepts to experiment with. While some of them perform very well, not all of them do so. In every system, we need to evaluate the contribution of the features and discard the unnecessary ones. This systematic way in which the components of a machine learning framework are typically deleted or replaced to determine how these changes affect the performance is called an ablation study. In this work, we performed a number of ablation tests to remove the redundant parts from the system.

*Ceiling Analysis*: To assess the upper-bound performance of the cascaded model, we conducted a ceiling analysis by providing ground truth ROIs directly to the segmentation model, bypassing the ROI extractor. This experiment allowed us to determine the ideal performance achievable by the segmentation model if the ROI detection stage were perfect, isolating the impact of ROI extraction errors on the overall cascaded system. According to the ceiling analysis, we obtained a Dice score, sensitivity, and mean IoU of 93.73%, 93.64%, and 86.66%, respectively, which is a significant boost compared to the cascaded model. This performance increase also highlights the segmentation model's potential when supplied with accurate ROI inputs and suggests that the current cascaded model's performance is partly constrained by inaccuracies in the ROI extractor.

*The cascaded model:* The cascaded model was a key factor in the excellent performance of the proposed system. We also tried our proposed segmentation model both with and without the ROI extraction model and found that the extracted ROI helped the model to improve its performance. Quantitively, using the cascaded network, the Dice score and the IoU increased to almost 7% and 10%, respectively.

*Number of decoder blocks:* Another critical factor was the number of decoder blocks. As the encoder and decoder blocks are connected with skip connections, an increase in decoder blocks also results in an increment in the number of encoder blocks, which may subsequently cause an overfitting change. We experimented with a number of different architectures varying the number of decoder and encoder blocks and found that the architecture works best if we use five decoder blocks.

*Preprocessing pipeline:* Preprocessing is an essential step in preparing the input data to enhance model performance. We experimented with different preprocessing techniques to evaluate their impact on the overall system. Specifically, we compared the performance of our model using raw images, with all preprocessing steps without augmentation and with all preprocessing steps with augmentation. The model can hardly perform well using raw images. The model performance is boosted significantly after including the heuristic slice removal, Gaussian blur, and CLAHE. Finally, the performance reaches its peak if we use augmentation techniques.

*Learning rate*: The learning rate stands as a pivotal hyperparameter in model training. Within our study, we explored various learning rates to ascertain the most optimized value for achieving superior performance alongside expedited training. Through the ablation study, we determined that learning rates of 0.1 and 0.01 were excessively large, impeding convergence. In contrast, a learning rate of 0.001 emerged as the optimized choice.

The results of the ablation studies are compiled in Table 8.

*Table 8: Result of the ablation study*

| Criterion | Value of the criterion | Dice score |
|---|---|---|
| Ceiling analysis | With ground truth ROI | **93.7%** |
| | With ROI from the detection model | 91% |
| The cascaded model | Without the ROI extractor model | 74.1% |
| | **With the ROI extractor model** | **84%** |
| Number of decoder blocks | 3 | 80% |
| | **5** | **84.2%** |
| | 7 | 82.3% |
| Preprocessing | No preprocessing | 68.3% |
| | With slice removal, blur, and CLAHE | 78.5% |
| | **With all the preprocessing steps** | **81.1%** |
| Learning rate | 0.0001 | 89.3% |
| | **0.001** | **89.7%** |
| | 0.01 | 84.1% |
| | 0.1 | 75.8% |

## 5. Conclusion

In this study, we have taken a machine-agnostic approach to automatically segment lumbar vertebrae and intervertebral discs from MR images using our proposed model. Traditional segmentation methods are often found to underperform in this task because of the extremely low contrast of the images and intra-scanner or inter-scanner variability of the produced images. Our solution achieves excellent performance by showing excellent robustness to segment different types of MR images and achieving a DSC, IoU, and pixel accuracy of 91.03%, 83.66%, and 99.78%, respectively. We hope that this approach will go a long way in automating diagnosis for a number of spine-related disorders and diseases.

Our study still has some limitations, including the lack of enough data, potential challenges in generalization to different modalities of images, and segmentation of other parts of the spine. To address these limitations, future research should focus on expanding and utilizing the dataset to ensure enough data samples and in-depth annotations. Additionally, the integration of real-time data collection and continuous model refinement in clinical practice holds promise for improving the accuracy and reliability of automated spine segmentation from MRI. Further exploration of interpretability and explainability techniques can also enhance the clinical utility of the model, providing insights into the decision-making process and increasing trust among healthcare providers. In conclusion, our study presents a promising avenue for advancing the field of MRI lumbar image segmentation with outstanding performance and robustness. We envision that our research will encourage further exploration in this domain, leading to enhanced diagnostic tools in the future.

**Funding:** This work was made possible by High Impact grant# QUHI-CENG-23/24-216 from Qatar University. The statements made herein are solely the responsibility of the authors. The open-access publication cost is covered by the Qatar National Library.

**Data Availability Statement**
The dataset utilized in this study can be obtained upon request from the corresponding author.

**Conflict of Interest**
The authors have no conflicts of interest to disclose for this study.

**References:**
Alsaleh, K., Ho, D., Rosas-Arellano, M. P., Stewart, T. C., Gurr, K. R., & Bailey, C. S. (2017). Radiographic assessment of degenerative lumbar spinal stenosis: is MRI superior to CT? *European Spine Journal, 26*, 362-367.

Aslan, M., Shalaby, A., Ali, A., & Farag, A. A. (2015). Model-based segmentation, reconstruction, and analysis of the vertebral body from spinal CT. *Spinal Imaging and Image Analysis*, 381-438.

Aslan, M. S., Ali, A., Rara, H., & Farag, A. A. (2010). An automated vertebra identification and segmentation in CT images. In *2010 IEEE international conference on image processing* (pp. 233-236): IEEE.

Bejnordi, B. E., Veta, M., Van Diest, P. J., Van Ginneken, B., Karssemeijer, N., Litjens, G., Van Der Laak, J. A., Hermsen, M., Manson, Q. F., & Balkenhol, M. (2017). Diagnostic assessment of deep learning algorithms for detection of lymph node metastases in women with breast cancer. *Jama, 318*, 2199-2210.

Bhole, C., Kompalli, S., & Chaudhary, V. (2009). Context sensitive labeling of spinal structure in MR images. In *Medical Imaging 2009: Computer-Aided Diagnosis* (Vol. 7260, pp. 1064-1072): SPIE.

Bogduk, N. (2005). *Clinical anatomy of the lumbar spine and sacrum*: Elsevier Health Sciences.

Chen, J., Lu, Y., Yu, Q., Luo, X., Adeli, E., Wang, Y., Lu, L., Yuille, A. L., & Zhou, Y. (2021). Transunet: Transformers make strong encoders for medical image segmentation. *arXiv preprint arXiv:2102.04306*.

Chen, L.-C., Papandreou, G., Schroff, F., & Adam, H. (2017). Rethinking atrous convolution for semantic image segmentation. *arXiv preprint arXiv:1706.05587*.

Corso, J. J., Alomari, R. S., & Chaudhary, V. (2008). Lumbar disc localization and labeling with a probabilistic model on both pixel and object features. In *Medical Image Computing and Computer-Assisted Intervention–MICCAI 2008: 11th International Conference, New York, NY, USA, September 6-10, 2008, Proceedings, Part I 11* (pp. 202-210): Springer.

Deans, S. R. (1981). Hough transform from the Radon transform. *IEEE transactions on pattern analysis and machine intelligence*, 185-188.

deci.ai. (2023). YOLO-NAS by Deci Achieves SOTA Performance on Object Detection Using Neural Architecture Search. In (Vol. 2023).

Deyo, R. A., Mirza, S. K., Martin, B. I., Kreuter, W., Goodman, D. C., & Jarvik, J. G. (2010). Trends, major medical complications, and charges associated with surgery for lumbar spinal stenosis in older adults. *Jama, 303*, 1259-1265.

Dolz, J., Desrosiers, C., & Ben Ayed, I. (2018). IVD-Net: Intervertebral disc localization and segmentation in MRI with a multi-modal UNet. In *International workshop and challenge on computational methods and clinical applications for spine imaging* (pp. 130-143): Springer.

Egger, J., Nimsky, C., & Chen, X. (2017). Vertebral body segmentation with GrowCut: Initial experience, workflow and practical application. *SAGE open medicine, 5*, 2050312117740984.

Elfwing, S., Uchibe, E., & Doya, K. (2018). Sigmoid-weighted linear units for neural network function approximation in reinforcement learning. *Neural Networks, 107*, 3-11.

Fayssoux, R., Goldfarb, N. I., Vaccaro, A. R., & Harrop, J. (2010). Indirect costs associated with surgery for low back pain—a secondary analysis of clinical trial data. *Population health management, 13*, 9-13.

Frost, B. A., Camarero-Espinosa, S., & Foster, E. J. (2019). Materials for the spine: anatomy, problems, and solutions. *Materials, 12*, 253.

Ghosh, S., & Chaudhary, V. (2014). Supervised methods for detection and segmentation of tissues in clinical lumbar MRI. *Computerized Medical Imaging and Graphics, 38*, 639-649.

Ghosh, S., Malgireddy, M. R., Chaudhary, V., & Dhillon, G. (2012). A new approach to automatic disc localization in clinical lumbar MRI: Combining machine learning with heuristics. In *2012 9th IEEE International Symposium on Biomedical Imaging (ISBI)* (pp. 114-117): IEEE.

Han, Z., Wei, B., Mercado, A., Leung, S., & Li, S. (2018). Spine-GAN: Semantic segmentation of multiple spinal structures. *Medical image analysis, 50*, 23-35.

Hoy, D., March, L., Brooks, P., Blyth, F., Woolf, A., Bain, C., Williams, G., Smith, E., Vos, T., & Barendregt, J. (2014). The global burden of low back pain: estimates from the Global Burden of Disease 2010 study. *Annals of the rheumatic diseases, 73*, 968-974.

Huang, G., Liu, Z., Van Der Maaten, L., & Weinberger, K. Q. (2017). Densely connected convolutional networks. In *Proceedings of the IEEE conference on computer vision and pattern recognition* (pp. 4700-4708).

Isensee, F., Petersen, J., Klein, A., Zimmerer, D., Jaeger, P. F., Kohl, S., Wasserthal, J., Koehler, G., Norajitra, T., & Wirkert, S. (2018). nnu-net: Self-adapting framework for u-net-based medical image segmentation. *arXiv preprint arXiv:1809.10486*.

Jarvik, J. G., & Deyo, R. A. (2002). Diagnostic evaluation of low back pain with emphasis on imaging. *Annals of internal medicine, 137*, 586-597.

Jeon, I., & Kong, E. (2020). Application of simultaneous 18F-FDG PET/MRI for evaluating residual lesion in pyogenic spine infection: A case report. *Infection & Chemotherapy, 52*, 626.

Jocher, G. (2022). ultralytics/yolov5: v7.0 - YOLOv5 SOTA Realtime Instance Segmentation. In (Vol. 2023): Zenodo.

Jocher, G., Chaurasia, A., & Qiu, J. (2023). YOLO by Ultralytics. In.

Khalil, Y. A., Becherucci, E. A., Kirschke, J. S., Karampinos, D. C., Breeuwer, M., Baum, T., & Sollmann, N. (2022). Multi-scanner and multi-modal lumbar vertebral body and intervertebral disc segmentation database. *Scientific Data, 9*, 97.

Kim, C., Park, S.-M., Lee, S., & Lee, D. (2024). A deep learning harmonization of multi-vendor MRI for robust intervertebral disc segmentation. *IEEE Access*.

Kiranyaz, S., Malik, J., Abdallah, H. B., Ince, T., Iosifidis, A., & Gabbouj, M. (2021). Self-organized operational neural networks with generative neurons. *Neural Networks, 140*, 294-308.

Lee, S., Lee, J. W., Yeom, J. S., Kim, K.-J., Kim, H.-J., Chung, S. K., & Kang, H. S. (2010). A practical MRI grading system for lumbar foraminal stenosis. *American Journal of Roentgenology, 194*, 1095-1098.

Li, C., Li, L., Geng, Y., Jiang, H., Cheng, M., Zhang, B., Ke, Z., Xu, X., & Chu, X. J. a. p. a. (2023). Yolov6 v3. 0: A full-scale reloading.

Li, C., Li, L., Jiang, H., Weng, K., Geng, Y., Li, L., Ke, Z., Li, Q., Cheng, M., & Nie, W. J. a. p. a. (2022). YOLOv6: A single-stage object detection framework for industrial applications.

Li, H., Xiong, P., An, J., & Wang, L. (2018). Pyramid attention network for semantic segmentation. *arXiv preprint arXiv:1805.10180*.

Liu, X., Deng, W., & Liu, Y. (2021). Application of hybrid network of UNet and feature pyramid network in spine segmentation. In *2021 IEEE International Symposium on Medical Measurements and Applications (MeMeA)* (pp. 1-6): IEEE.

Lootus, M., Kadir, T., & Zisserman, A. (2014). Vertebrae detection and labelling in lumbar MR images. In *Computational Methods and Clinical Applications for Spine Imaging: Proceedings of the Workshop held at the 16th International Conference on Medical Image Computing and Computer Assisted Intervention, September 22-26, 2013, Nagoya, Japan* (pp. 219-230): Springer.

Lu, H., Li, M., Yu, K., Zhang, Y., & Yu, L. (2023). Lumbar spine segmentation method based on deep learning. *Journal of Applied Clinical Medical Physics*, e13996.

Naegel, B. (2007). Using mathematical morphology for the anatomical labeling of vertebrae from 3D CT-scan images. *Computerized Medical Imaging and Graphics, 31*, 141-156.

Oktay, A. B., & Akgul, Y. S. (2013). Simultaneous localization of lumbar vertebrae and intervertebral discs with SVM-based MRF. *IEEE transactions on biomedical engineering, 60*, 2375-2383.

Oktay, O., Schlemper, J., Folgoc, L. L., Lee, M., Heinrich, M., Misawa, K., Mori, K., McDonagh, S., Hammerla, N. Y., & Kainz, B. (2018). Attention u-net: Learning where to look for the pancreas. *arXiv preprint arXiv:1804.03999*.

Redmon, J., Divvala, S., Girshick, R., & Farhadi, A. (2016). You only look once: Unified, real-time object detection. In *Proceedings of the IEEE conference on computer vision and pattern recognition* (pp. 779-788).

Reza, A. M. (2004). Realization of the contrast limited adaptive histogram equalization (CLAHE) for real-time image enhancement. *Journal of VLSI signal processing systems for signal, image and video technology, 38*, 35-44.

Ronneberger, O., Fischer, P., & Brox, T. (2015). U-net: Convolutional networks for biomedical image segmentation. In *Medical Image Computing and Computer-Assisted Intervention–MICCAI 2015: 18th International Conference, Munich, Germany, October 5-9, 2015, Proceedings, Part III 18* (pp. 234-241): Springer.

Sneath, R. J., Khan, A., & Hutchinson, C. (2022). An Objective Assessment of Lumbar Spine Degeneration/Ageing Seen on MRI Using An Ensemble Method—A Novel Approach to Lumbar MRI Reporting. *Spine, 47*, E187-E195.

Wang, S., Jiang, Z., Yang, H., Li, X., & Yang, Z. (2022). Automatic Segmentation of Lumbar Spine MRI Images Based on Improved Attention U-Net. *Computational Intelligence and Neuroscience, 2022*.

Wang, Z., Xiao, P., & Tan, H. (2023). Spinal magnetic resonance image segmentation based on U-net. *Journal of Radiation Research and Applied Sciences, 16*, 100627.

Warfield, S. K., Zou, K. H., & Wells, W. M. (2004). Simultaneous truth and performance level estimation (STAPLE): an algorithm for the validation of image segmentation. *IEEE transactions on medical imaging, 23*, 903-921.

Westbrook, C., & Talbot, J. (2018). *MRI in Practice*: John Wiley & Sons.

Whitehead, W., Moran, S., Gaonkar, B., Macyszyn, L., & Iyer, S. (2018). A deep learning approach to spine segmentation using a feed-forward chain of pixel-wise convolutional networks. In *2018 IEEE 15th International Symposium on Biomedical Imaging (ISBI 2018)* (pp. 868-871): IEEE.

Wittens, M. M. J., Allemeersch, G.-J., Sima, D. M., Naeyaert, M., Vanderhasselt, T., Vanbinst, A.-M., Buls, N., De Brucker, Y., Raeymaekers, H., & Fransen, E. (2021). Inter-and intra-scanner variability of automated brain volumetry on three magnetic resonance imaging systems in Alzheimer's disease and controls. *Frontiers in Aging Neuroscience, 13*, 746982.

Yilizati-Yilihamu, E. E., Yang, J., Yang, Z., Rong, F., & Feng, S. (2023). A spine segmentation method based on scene aware fusion network. *BMC neuroscience, 24*, 49.

Zhang, Q., Du, Y., Wei, Z., Liu, H., Yang, X., & Zhao, D. (2021). Spine medical image segmentation based on deep learning. *Journal of Healthcare Engineering, 2021*.

Zhang, R., Xiao, X., Liu, Z., Li, Y., & Li, S. (2020). MRLN: Multi-task relational learning network for mri vertebral localization, identification, and segmentation. *IEEE journal of biomedical and health informatics, 24*, 2902-2911.

Zhao, H., Shi, J., Qi, X., Wang, X., & Jia, J. (2017). Pyramid scene parsing network. In *Proceedings of the IEEE conference on computer vision and pattern recognition* (pp. 2881-2890).

Zhou, J., Damasceno, P. F., Chachad, R., Cheung, J. R., Ballatori, A., Lotz, J. C., Lazar, A. A., Link, T. M., Fields, A. J., & Krug, R. (2020). Automatic vertebral body segmentation based on deep learning of Dixon images for bone marrow fat fraction quantification. *Frontiers in Endocrinology, 11*, 612.

Zhou, Z., Siddiquee, M. M. R., Tajbakhsh, N., & Liang, J. (2019). Unet++: Redesigning skip connections to exploit multiscale features in image segmentation. *IEEE transactions on medical imaging, 39*, 1856-1867.